\documentclass[a4paper,12pt]{article}

\usepackage{amssymb,epsf,graphics,amsmath}

\newtheorem{theo}{Theorem}[section]
\newtheorem{prop}[theo]{Proposition}

\newtheorem{lem}[theo]{Lemma}

\newenvironment{proof}{{\bf Proof }}{\hfill $\Box$}

\newcommand{\CC}{\mathbb{C}}
\newcommand{\NN}{\mathbb{N}}
\newcommand{\RR}{\mathbb{R}}

\newcommand{\rA}{\mathcal{A}}

\newcommand{\rH}{\mathcal{H}}

\newcommand{\rN}{\mathcal{N}}
\newcommand{\rM}{\mathcal{M}}

\newcommand{\rP}{\mathcal{P}}

\newcommand{\TF}{T\Phi}
\newcommand{\s}{\sigma}
\font\timesept=cmr7

\begin{document}


\title{Repeated Quantum Interactions\\and\\Unitary Random Walks}

\author{St\'ephane ATTAL${}^{{}^1}$ and  Ameur DHAHRI${}^{{}^2}$}

\date{}

\maketitle

\centerline{\timesept ${}^{{}^1}$ Universit\'e de Lyon, Universit\'e de Lyon 1}
\vskip -1mm
\centerline{\timesept Institut Camille Jordan, U.M.RM 5208}
\vskip -1mm
\centerline{\timesept 21 av Claude Bernard}
\vskip -1mm
\centerline{\timesept 69622 Villeubanne cedex, France}

\bigskip
\centerline{\timesept ${}^{{}^2}$ CEREMADE}
\vskip -1mm
\centerline{\timesept Universit\'e de Paris-Dauphine}
\vskip -1mm
\centerline{\timesept Place du Mar\'echal De Lattre De Tassigny}
\vskip -1mm
\centerline{\timesept 75775 Paris cedex 16, France}

\bigskip

\begin{abstract}
Among the discrete evolution equations describing a quantum system
$\rH_S$ undergoing repeated quantum
interactions with a chain of exterior systems, we study and
characterize those which are directed by classical random
variables in
$\RR^N$. The characterization we obtain is entirely algebraical in terms of the
unitary operator driving the elementary interaction. We show that the
solutions of these equations are then random walks on the group $U(\rH_0)$
of unitary operators on $\rH_0$.
\end{abstract}

\section{Introduction}
In the article \cite{AP} Attal and Pautrat have explored the
Hamiltonian description of a quantum system undergoing repeated
interactions with a chain of quantum systems. They have shown that these
``deterministic'' dynamics give rise to  quantum stochastic
differential equations in the continuous limit. Since that result, some
interest has been found in the repeated quantum interaction model in
itself (cf \cite{AJ1}, \cite{AJ2}, \cite{BJM1}, \cite{BJM2}, \cite{BJM3})
and several physical works are
in progress on that subject (for example \cite{AKP}). These repeated
interaction models are interesting for several reasons:

\smallskip\noindent
-- they provide a quantum dynamics which is at the same time
   Hamiltonian and Markovian,

\smallskip\noindent
-- they allow to implement easily the dissipation for a quantum
   system, in particular they are practical models for simulation.

\smallskip\noindent
The probabilistic nature of the continuous limit found by Attal and
Pautrat is not due to the passage to the limit, it is already built-in
the Hamiltonian dynamics of repeated quantum interactions (it is
actually built-in the axioms of quantum mechanics). 

The evolution equations decribing the repeated quantum interactions
are purely deterministic but they already show up terms which can be
interpreted as ``discrete time quantum noises''. The point with these
discrete quantum noises is that sometimes they may give rise to
classical noises. That is, some linear combinations of these quantum
noises happen to be mutually commuting families of Hermitian operators, hence
they simultaneously diagonalize and they can be represented as
classical stochastic processes.

In the other cases, that is, with different combinations of the
quantum noises, no classical process emerges and the dynamics of
repeated quantum interactions is purely quantum.

The aim of the article is to explore the case when the dynamics is
classically driven. We characterise algebraically, on the Hamiltonian,
the case when the dynamics is classical.

The article is organised as follows. We first (Section \ref{RQI})
present the physical and mathematical setups for the repeated quantum
interactions. In Section \ref{CRW} we introduce the basic algebric
tool: the obtuse random walk which are an appropriate ``basis'' of
random walks adapted to this language. We then explore and
characterise the unitary random walks which emerge classically from
the repeated quantum interactions (Section \ref{RWU}). We finally 
specialize in Section \ref{N=1} our result to the one dimensional case which alreay shows
up a non-trivial structure.

\section{Repeated Quantum Interactions}\label{RQI}

\subsection{The Physical Model}\label{RQIPhys}

Repeated quantum interaction models  are  physical models developed by
Attal and Pautrat in \cite{AP} which consist in describing the Hamiltonian
dynamics of a quantum system undergoing a sequence of interactions
with an  environment made of a chain of identical systems. These
models were developed for they furnish a toy model for a quantum
dissipative system, they are at the same time Hamiltonian and
Markovian, they spontaneously give rise to quantum stochastic
differential equations in the continuous time limit.  Let us describe
precisely the physical and the mathematical setup of these models.

\smallskip
We consider a reference quantum system with state space $\rH_0$, which
we shall call the \emph{small system} (even if it is not that
small!). Another system $\rH_E$, called the \emph{environment} is made up of a
chain of identical copies of a quantum system $\rH$, that is,
$$
\rH_E=\bigotimes_{n\in\NN^*} \rH
$$
where the countable tensor product is understood in a sense that we
shall make precise later. 

The dynamics in between $\rH_0$ and $\rH_E$ is driven as follows. The
small system $\rH_0$ interacts with the first copy $\rH$ of the chain
during an interval $[0,h]$ of time and following an Hamiltonian $H$ on
$\rH_0\otimes\rH$. That is, the two systems evolve together following
the unitary operator 
$$
U=e^{-ihH}\,.
$$
After this first interaction, the small system $\rH_0$ stops
interacting with the first copy and starts an interaction with the
second copy which was left unchanged until then. This second
interaction follows the same unitary operator $U$. And so on, the
small system $\rH_0$ interacts repeatedly with the elements of the
chain one after the other, following the same unitary evolution $U$. 

\smallskip
Let us give a mathematical setup to this repeated quantum interaction
model.

\subsection{The Mathematical Setup}\label{S:RQIMath}

Let $\rH_0$ and $\rH$ be two separable Hilbert spaces (in the
following, for our probabilistic interpretations the space $\rH$ will
be choosen to be finite dimensional). We choose a fixed orthonormal
basis $\{X^n; n\in\rN\cup\{0\}\}$ where $\rN=\NN^*$ or $\{1,\ldots,
N\}$ depending on wether $\rH$ is infinite dimensional or not (note
the particular role played by the vector $X^0$ in our notation). We
consider the Hilbert space
$$
\TF=\bigotimes_{n\in\NN^*}\rH
$$
where this countable tensor product is understood with respect to the
stabilizing sequence $(X^0)_{n\in\NN^*}$. This is to say that an
orthonormal basis of $\TF$ is made of the vectors 
$$
X_\s=\bigotimes_{n\in\NN^*}X_{n}^{i_n}
$$
where $\s=(i_n)_{n\in\NN^*}$ runs over the set $\rP$ of all sequences in
$\rN\cap\{0\}$ with only a finite number of terms  different
of 0. 

\smallskip
Let $U$ be a fixed unitary operator on $\rH_0\otimes\rH$. We denote by
$U_n$ the natural ampliation of $U$ to $\rH_0\otimes\TF$ where $U_n$
acts as $U$ on the tensor product of $\rH_0$ and the $n$-th copy of
$\rH$ and $U$ acts as the identity of the other copies of $\rH$. In
our physical model, the operator $U_n$ is the unitary operator
expressing the result of the $n$-th interaction. We also define
$$
V_n=U_n\,U_{n-1}\ldots U_1,
$$
with the convention $V_0=I$. Physically, $V_n$ is clearly the unitary
operator expressing the transformation of the whole system after the
$n$ first interactions. 

\smallskip
Define the elementary operators $a^i_j$, $i,j\in\rN\cap\{0\}$ on $\rH$
by
$$
a^i_j X^k=\delta_{i,k}X^j\,.
$$
We denote by $a^i_j(n)$ their natural ampliation to $\TF$ acting on
the $n$-th copy of $\rH$ only. That is, if $\s=(i_n)_{n\in\NN^*}$
$$
a^i_j(n)X_\s=\delta_{i,i(n)}X_{\s\setminus\{i_n\}\cup\{j_n\}}\,.
$$
One can easily prove (in the finite dimensional case  this is obvious,
in the infinite dimensional case it is an exercise) that $U$ can
always be written as
$$
U=\sum_{i,j\in\rN\cup\{0\}} U^i_j\otimes a^i_j
$$
for some bounded operators $U^i_j$ on $\rH_0$ such that:

\smallskip\noindent
-- the series above is strongly convergent,

\smallskip\noindent
-- $\sum_{k\in\rN\cup\{0\}} (U^k_i)^*\, U^k_j=\sum_{k\in\rN\cup\{0\}}  U^k_j\,(U^k_i)^*=\delta_{i,j}I$.

\smallskip
With this representation for $U$, it is clear that the operator $U_n$,
representing the $n$-th interaction, is given by
$$
U_n=\sum_{i,j\in\rN\cup\{0\}} U^i_j\otimes a^i_j(n)\,.
$$
With these notations, the sequence ${(V_n)}$ of
unitary operators describing the $n$ first repeated interactions can be
represented as follows:
\begin{align*}
V_{n+1}&=U_{n+1}\, V_n\\
&=\sum_{i,j\in\rN\cup\{0\}} U^i_j\otimes a^i_j(n+1)V_n\,.
\end{align*}
But, inductively, the operator $V_n$ acts only on the $n$ first sites
of the chain $\TF$, whereas the operators $a^i_j(n+1)$ acts on the
$(n+1)$-th site only. hence they commute. In the following, we shall
drop the $\otimes$ symbols, identifying operators like $a^i_j(n+1)$
with $I_{\rH_0}\otimes a^i_j(n+1)$. This gives finally
\begin{equation}\label{E:Vn}
V_{n+1}=\sum_{i,j\in\rN\cup\{0\}} U^i_j\,V_n a^i_j(n+1)\,.
\end{equation}

In Quantum Probability Theory, the operators $a^i_j(n)$ have a
particular interpretation, they are \emph{discrete-time quantum
noises}, they describe the different types of basic innovations than
can be brought by the environment when interacting with the small
system. See \cite{At} for complete details on that theory, the
understanding of which is not necessary here.

The only important point to understand at that stage is the
following. In some cases the above equation (\ref{E:Vn}) corresponds
to an equation divren by a \emph{classical noise}, i.e. driven by a
\emph{random walk}. This is what we shall describe in the next
section.

\section{Classical Random Walks}\label{CRW}

In order to understand the link that may exist between the
discrete-time quantum noises $a^i_j$ and classical random walk, one
needs to pass through a particular family of random walks, the
\emph{obtuse random walks}. Defined by  Attal and Emery in \cite{A-E},
these random walks constitute a kind of basis of all the random walks
in $\RR^N$. Let us describe them.

\subsection{Obtuse Random Walks in $\RR^N$}

Let $X$ be a random variable in $\RR^N$ taking $N+1$ values
$v_0,...,v_N$ with respective probabilities $p_0,...,p_N$ such that
$p_i>0,\,\forall i\in\{0,1,...,N\}$. The canonical space of $X$ is
the triple $(A,\,\mathcal{A},\,P)$, where $A=\{0,1,...,N\}$,
$\mathcal{A}$ is the $\sigma-$field of subsets of $A$ and $P$ is the
probability measure given by $P(\{i\})=p_i$. Hence for all
$i\in\{0,1,...,N\}$ we have $X(i)=v_i$ and $P(X=v_i)=P(\{i\})=p_i$. 

\smallskip
We say that such a random variable  $X$ is \emph{centered} if
$\mathbb{E}(X)=0$ (as a vector of $\RR^N$). We say that $X$ is
\emph{normalized} if $Cov(X)=I$ (as a $N\times N$-matrix). 

\smallskip
Let us denote by $X^1,...,X^N$ the coordinates of $X$ in the canonical
basis of $\RR^N$ and define the random variable $X^0$
on $(A,\,\mathcal{A},\,P)$ given by $X^0(i)=1,\;\forall i\in A$. Let
us introduce the random variables $\widetilde{X}^i$ defined by
$$
\widetilde{X}^i(j)=\sqrt{p_j}\, X^i(j)\,,
$$
for all $
i,\,j\in\{0,1,...,N\}$. We then have the following easy
characterization (cf \cite{A-E}).  
\begin{prop}\label{P:centered}
The following assertions are equivalent:
\begin{enumerate}
\item[1)] The random variable $X$ is centered and normalized,
\item[2)] The family $v_0,...,v_N$ of values of $X$ satisfies
$<v_i\,,\,v_j>=-1$, for all $i\neq j$ and the probabilities $p_i$'s are
given by  
$$
p_i=\frac{1}{1+||v_i||^2},\ \ \mbox{for all } i\in\{0,1,...,N\}\,,
$$
\item[3)] The matrix $(\widetilde{X}^0,\widetilde{X}^1,...,\widetilde{X}^N)$ is unitary.
\end{enumerate}
\end{prop}

A family of $N+1$ vectors in $\RR^N$ satisfying the above condition
$$
<v_i,\,v_j>=-1\,,
$$
for all $i\not = j$, is called an \emph{obtuse system} in
\cite{A-E}. Hence, a random variable $X$ satisfying one of the above
condition is called an \emph{obtuse random variable}. 

\smallskip
Note that, as a corollary of the above proposition, the random variables
$X^0,X^1,...,X^N$ are linearly independent and hence they form an
orthonormal basis of
$L^2(A,\,\mathcal{A},\,P)$. 
In particular, for every $i,j\in\{1,\ldots,N\}$ the random variable
$X^iX^j$ can be decomposed into
$$
X^iX^j=\sum_{k=0}^N\,T_k^{ij}X^k
$$
for some real coefficients $T^{ij}_k$. The familly of such
coefficients forms a so-called 3-tensor, that is they are the
coordinates of a linear mapping $T$ from $\RR^N$ to $M_N(\RR)$.

 We say that a 3-tensor $T$ is
\emph{sesqui-symmetric} if the two following assumptions are satisfied: 
\begin{enumerate}
\item[i)] $(i,\,j,\,k)\longmapsto T_k^{ij}$ is symmetric, 
\item[ii)] $(i,\,j,\,l,\,m)\longmapsto\sum_{k=1}^N\, T_k^{ij}
T_k^{lm}+\delta_{ij}\delta_{lm}$ is symmetric. 
\end{enumerate}

Using the commutativity and the associativity of the product  $X^iX^j$
it is easy to prove the following (cf \cite{A-E}). 

\begin{theo}
If $X$ is a centered and normalized random variable in $\RR^N$, taking
exactly $N+1$ values, then there exists a sesqui-symmetric 3-tensor
$T$ such that 
\begin{eqnarray*}\label{representation}
X\otimes X=I+T(X)\,.
\end{eqnarray*}
\end{theo}

In the following, by an \emph{obtuse random walk} we mean a sequence
${(X_p)}_{p\in\NN}$ of independent copies of a given obtuse random
variable $X$. Actually, the random walk is the sequence made of the
partial sums $\sum_{p\leq n} X_p$, but we shall not make any
disctinctions between the two processes in the terminology.

\subsection{More General Random Variables}

We claimed above that obtuse random variables are a kind of basis for
the random variables in $\RR^N$ in general. Let us make precise here
what we mean by that.

First of all, a remark on the number $N+1$ of values attached to $X$
in $\RR^N$. If one had asked that $X$ takes less than $N+1$ values in
$\RR^N$ ($k$, say)and be centered and normalized too, it is not difficult to
show that $X$ is actually taking values on a proper subspace of
$\RR^N$, with dimension $k-1$. For example, a centered, normalized
random variable in $\RR^2$ which takes only two different values, is
living on a line.

Now, if $Y$ is a random variable in $\RR^N$ taking $k$ different
possible values $w_1, \ldots,w_k$, with probability $p_1, \ldots, p_k$
and $k>n+1$. Consider an obtuse random variable $X$ in $\RR^{k-1}$
taking values $v_1, \ldots, v_k$ with the same probabilities $p_1,
\ldots, p_k$ as those of $Y$. We have seen that the coordinate random variables $X^1,
\ldots, X^{k-1}$, together with the deterministic random variable
$X^0$, form an orthonormal basis of $L^2(A,\rA,P)$ we can represent
each of the coordinates of $Y$ as
$$
Y^i=\sum_{j=0}^{k-1} \alpha^i_j\, X^j\,.
$$
Hence we have a simple representation of $Y$ in terms of a given
obtuse random variable $X$.

\subsection{Connecting With the Discrete Quantum Noises}

The obtuse random walks admit a very simple and natural representation
in terms of the operators $a^i_j(n)$ defined in Section
\ref{S:RQIMath}. 

\smallskip
Let $X$ be an obtuse random variable in
$\RR^N$. On the product space
$(A^\NN,\,\mathcal{A}^{\otimes\NN},\,P^{\otimes\NN})$ we define a
sequence $(X_p)_{p\in\NN}$ of independent, identically distributed,
random variables, each with the same law as $X$. 

Consider the space 
$T\Phi(X)=L^2(A^\NN,\,\mathcal{A}^{\otimes\NN},\,P^{\otimes\NN})$ and the
random variables
$$
X_A=\prod_{(p,i)\in A}X^i(p)\,,
$$ 
where $A$ is any sequence in  $\{0,1,...,N\}$ with only finitely many
terms different from 0. 

The following result is also easy to prove (cf \cite{At}).

\begin{prop}\label{P:base}
The random variables $X_A$, where $A$ runs over the sequences in  $\{0,1,...,N\}$ with only finitely many
terms different from 0, form an orthonormal basis of $T\Phi(X)$.
\end{prop}

In particular we see that there exists a very natural Hilbert space
isomorphism between the space $T\Phi(X)$ and the chain $\TF$
constructed in Section \ref{S:RQIMath}, over the space
$\rH=\CC^{N+1}$. Regarding this isosmorphism, one can consider the
operator $\rM_{X^i(p)}$ of multiplication by the random variable
$X^i_p$ on $T\Phi(X)$. This self-adjoint operator contains all the
probabilistic information associated to the random variable $X^i_p$,
it admits the same functional calculus, etc ... it is the actual
representant of the random variable $X^i_p$ in this Hilbert space
setup.

As each of the probabilistic space $T\Phi(X)$ are made isomorphic to
$\TF$ we can naturally wonder what happens to the operators
$\rM_{X^i(p)}$ through this identification. The answer is surprisingly
simple (cf \cite{At}).

\begin{theo}\label{identi}
Let $X$ be an obtuse random variable in $\RR^N$ and let
$(X_p)_{p\in\NN}$ be the associated random walk on the canonical space
$T\Phi(X)$. Let $T$ be the sesqui-symmetric 3-tensor associated to
$X$. If we denote by $U$ the natural unitary isomorphism from
$T\Phi(X)$ to $T\Phi$ , then for all $p\in\NN,\;i\in\{1,...,N\}$ we
have 
$$U\rM_{X^i_p}U^*=a^0_i(p)+a_0^i(p)+\sum_{j,l=1}^N\,T_i^{jl}a_l^j(p)\,.$$
\end{theo}

Here we are! By a simple linear combination of the basic matrices
$a^i_j(p)$ one can reproduce any random variable on $\RR^N$. 

Coming back to the evolution equation (\ref{E:Vn}), we see basically
two different cases may appear. 

First case: the coefficients $U^i_j$ of the basic unitary matrix $U$
are such that the equation \ref{E:Vn} reduces to something like
$$
V_{n+1}=AV_n+\sum_{i=1}^N B_iV_n\rM_{X^i_p}\,.
$$
This means that this operator-valued evolution equation, when
transported back to $T\Phi(X)$ is an operator-valued (actually unitary
operator-valued) equation driven
by a random walk ${(X_p)}_{p\in\NN}$. It is a random walk on $U(N)$.

Second case: there is no such arrangement in the equation \ref{E:Vn},
this means it is purely quantum, it cannot be expressed via classical
noises, only quantum noises.

Our aim, in the rest of the article is to characterize completely
those unitary operators $U$ which give rise to a classically driven
evolution (first case).

\section{Random Walks on $U(\rH_0)$}\label{RWU}

In this section we work on the state space 
$$
\TF=\bigotimes_{n\in\NN^*}\CC^{N+1}\,.
$$
We consider a fixed obtuse random variable $X$, with values
$v_1,\ldots, v_n$ and with associed 3-tensor
$T$. We identify the operator
$$
a^0_i(p)+a_0^i(p)+\sum_{j,l=1}^N\,T_i^{jl}a_l^j(p)
$$ 
with the random variable $X^i_p$ and we denote it by $X^i_p$, instead
of $\rM_{X^i_p}$. Recall that $X^0_p$ is the constant random variable
equal to 1, hence as a multiplication operator on $\TF$ it coincides
with the identity operator $I$. 

In the following we extend the coefficients of the 3-tensor $T$ to the
set $\{0, 1,\ldots, N\}$. This extension is achieved by assigning the
following values:
$$
T^{ij}_0=T^{i0}_j=T^{0i}_j=\delta_{i,j}\,.
$$
With that extension, the second sesqui-symmetric relation for $T$ is written
simply 
$$
ii)\ \ \ (i,\,j,\,l,\,m)\longmapsto\sum_{k=0}^N\, T_k^{ij} T_k^{lm}\ \
\mbox{is symmetric.}
$$
 
Recall the discrete time evolution equation (\ref{E:Vn}) associated to
the repeated quantum interactions:
$$
V_{n+1}=\sum_{i,\,j=0}^N\,U_j^iV_na_j^i(n+1)\,,
$$
with the convention $
V_0=I$.

\begin{prop} The  discrete-time  evolution equation (\ref{E:Vn}) can be written as  
$$
V_{n+1}=\sum_{i=0}^N B_i\,V_nX^i_{n+1}\,,
$$
for some operators  $B_k$ on $\rH_0$,
if and only if the coefficients $U^i_j$ are of the form 
\begin{equation}\label{E:Uij1}
U^i_j=\sum_{k=0}^N\,T_k^{ij}B_k\,.
\end{equation}
\end{prop}
\begin{proof}
Let us prove first the sufficient direction. If $U$ is of the form
(\ref{E:Uij1}) then
\begin{eqnarray*}
V_{n+1}&=&\sum_{i,j=0}^N\,U_j^iV_na_j^i(n+1)\\
      &=&U^0_0V_n\,a_0^0(n+1)+\sum_{i=1}^N\,U_0^iV_n\,a^i_0(n+1)+
\sum_{i=1}^N\,U^0_iV_n\,a_i^0(n+1)+\\
&&\ \ +\sum_{i,j=1}^N\,U^i_jV_n\,a^i_j(n+1)\,.  
\end{eqnarray*}
The relation (\ref{E:Uij1}) implies in particular
$U^0_0=B_0$ and $U_i^0=U^i_0=B_i$.
This gives
\begin{eqnarray*}
V_{n+1}&=&B_0V_n\,
a^0_0(n+1)+\sum_{i=1}^N\,B_iV_n\,(a^i_0(n+1)+a_i^0(n+1))+\\
&&\ \ +\sum_{k=1}^N\,\sum_{i,j=1}^N T^{ij}_k B_kV_n\,
a^i_j(n+1)+\sum_{i=0}^N\,B_0V_n\,a_i^i(n+1)\\
   &=&B_0V_n+\sum_{k=1}^N\,B_kV_n\,\bigl[a_0^i(n+1)+a_i^0(n+1)+\sum_{i,j=1}^N\,T_k^{ij}a^i_j(n+1)\bigr]\\
      &=&B_0V_n+\sum_{k=1}^N\,B_kV_n\,X^k_{n+1}\\
      &=&\sum_{k=0}^N\,B_kV_n\,X^k_{n+1}\,.
\end{eqnarray*}
This gives the requiered result in one direction. The converse is easy
to prove by reversing all the arguments above.
\end{proof}

\smallskip
Now, consider the operators
$$
W_l=\sum_{i=0}^N\,v_i^lB_i\,,$$
with the convention $v_k^0=1,$ for all $k\in\{0,1,...,N\}$.  Our
purpose in the sequel is to prove that these operators are unitary if
and only if the evolution operator $U$ is unitary. Here is the first
step.

\begin{prop}\label{first sens}
If $U$ a unitary operator, then for all $l\in\{0,1,...,N\}$ the operator $W_l$ is unitary.
\end{prop}
\begin{proof}
We have
$$
W_lW_l^*=\sum_{i,j=0}^N\,v_l^iv_l^jB_iB_j^*\,.
$$
But expressing coordinate-wise the relation (\ref{representation}), we
have
$$v_l^iv_l^j=\sum_{m=0}^N\,T^{ij}_mv_l^m\,.$$
Hence, we get
\begin{eqnarray*}
W_lW_l^*&=&\sum_{i,j,m=0}^N\,T^{ij}_mv_l^mB_iB_j^*\\
        &=&\sum_{j,m=0}^N\,v_l^m\bigl(\sum_{i=0}^N\,T_m^{ij}B_i\bigr)B_j^*\\
&=&\sum_{j,m=0}^N\,v_l^mU^j_mU_j^{0*}\\
       &=&\sum_{m=0}^N\,v_l^m\big(\sum_{j=0}^N\,U_m^jU_j^{0*}\bigr)\\
&=&\sum_{m=0}^N\,v_l^m\big(\sum_{j=0}^N\,\delta_{m0}I\bigr)\\
&=&v_l^0I=I\,.
\end{eqnarray*}
This completes the proof.
\end{proof}

\smallskip
Now, our aim is to prove the converse of Proposition \ref{first
sens}. In order to achieve this, we need to express the coefficients
$U^i_j$ of $U$  in terms of the operators $W_l$'s. This is the aim of
the  following two lemmas. 

\begin{lem}\label{lemm}
For all $i\in\{0,1,...,N\}$ we have
\begin{eqnarray*}
B_i=\sum_{l=0}^N\,p_l\,v_l^i\,W_l\,.
\end{eqnarray*}
\end{lem}
\begin{proof}
We have
\begin{eqnarray*}
\sum_{l=0}^N\,p_l\,v_l^i\,W_l&=&\sum_{l=0}^N\,p_l\,v^i_l\,\big(\sum_{j=0}^N\,v_l^j\,B_j\bigr)\\
                        &=&\sum_{j=0}^N\,B_j\big(\sum_{l=0}^N\,p_l\,v_l^iv_l^j\bigr)\\
                        &=&\sum_{j=0}^N\,B_j\,\mathbb{E}(X^iX^j)\\
                        &=&\sum_{j=0}^N\,B_j\delta_{ij}=B_i\,.
\end{eqnarray*}
This ends the proof. 
\end{proof}
\begin{lem}\label{coeff-random}
For all $l,\;k\in\{0,1,...,N\}$ we have
\begin{eqnarray*}
U^k_l=\sum_{i=0}^N\,p_i\,v_i^kv_i^l\,W_i\,.
\end{eqnarray*}
\end{lem}
\begin{proof}
Recall that we have 
$$U^k_l=\sum_{j=0}^N\,T_j^{kl}B_j$$
and 
\begin{equation}\label{v}
v_i^lv_i^k=\sum_{j=0}^NT_j^{kl}v_i^j.
\end{equation}
By using Lemma \ref{lemm} and relation (\ref{v}), we get
\begin{eqnarray*}
U^k_l&=&\sum_{i,j=0}^N\,p_i\,T_j^{kl}\,v_i^j\,W_i\\
     &=&\sum_{i=0}^N\,p_i\,W_i\bigl(\sum_{j=0}^N\,T_j^{kl}\,v_i^j\bigr)\\
     &=&\sum_{i=0}^N\,p_i\,v_i^kv_i^l\,W_i.
\end{eqnarray*}
\end{proof}

As a corollary of the two above lemmas, we prove the following.
\begin{prop}\label{recip}
If all the operators $W_i$, for $i\in\{0,1,...,N\}$, are unitary  then the operator $U$ is unitary.
\end{prop}
\begin{proof}
We have
\begin{eqnarray*}
\sum_{k=0}^N\,(U_k^l)(U_m^k)^*&=&\sum_{i,j,k=0}\,p_ip_j\,v_i^kv_j^kv_i^lv_j^m\,W_iW_j^*\\
                            &=&\sum_{i,k=0}^Np_i^2\,(v_i^k)^2v_i^lv_i^m\,I+\sum_{i,j,k=0,\,i\neq
                            j}^N\,p_ip_j\,v_i^kv_j^kv_i^lv_j^m\,W_iW_j^*\\ 
                            &=&\sum_{i=0}^N\,p_i\bigl(p_i(||v_i||^2+1)\bigr)v_i^lv_i^m\,I+\\
&&\ \ \ +\sum_{i,j=0,\,i\neq j}^N\,p_ip_j\bigl(\sum_{k=0}^N\,v_i^kv_j^k\bigr)v_i^lv_j^m\,W_iW_j^*\\
                            &=&\sum_{i=0}^N\,p_i\bigl(p_i(||v_i||^2+1)\bigr)v_i^lv_i^m\,I+\\
&&\ \ \ +\sum_{i,j=0,i\neq j}^N\,p_ip_j(<v_i,\,v_j>+1)v_i^lv_j^m\,W_iW_j^*\,.
\end{eqnarray*}
But recall that, by Proposition \ref{P:centered},  we have
$p_i(||v_i||^2+1)=1$ and   $<v_i,\,v_j>=-1$ for all $i\neq j$. Therefore we get
$$\sum_{k=0}^N\,(U_k^l)(U_m^k)^*=\mathbb{E}(X^lX^m)I=\delta_{ml}I\,.$$
We have prove the unitary character of $U*$.
\end{proof}

Alltogether  we have proved the following result, which resumes all
the results obtained above.

\begin{theo}\label{iff}
Let $X$ be an obtuse random walk in $\RR^N$, with values $v_0, \ldots,
v_N$, with probabilities $P_0, \ldots, p_N$ and with 3-tensor $T$. Let
${(X_p)}_{p\in\NN}$ be its associated obtuse random walk.
Then the repeated quantum interaction evolution equation
$$
V_{n+1}=\sum_{i,j=0}^N U^i_jV_n\, a^i_j(n+1)
$$
takes the form
$$
V_{n+1}=\sum_{k=0}^N B_k V_n X^k_{n+1}
$$
if and only if there exists unitary operators $W_i$,
$i\in\{0,\ldots,N\}$,  on $\rH_0$ such
that the coefficients $U^i_j$ of $U$ are of the form
$$
U^k_l=\sum_{i=0}^N\,p_i\,v_i^kv_i^l\,W_i\,.
$$
In that case, the coefficients $B_k$ above are given by
$$
B_k=\sum_{l=0}^N\,p_l\,v_l^k\,W_l\,.
$$
\end{theo}

When the conditions above are satisfied, the evolution equation
$$
V_{n+1}=\sum_{k=0}^N B_k V_n X^k_{n+1}
$$
is, when seen in the space $T\Phi(X)$, an operator-valued evolution equation, driven by a
random walk. It is natural to wonder what king of stochastic process
it gives rise to.

\begin{theo}\label{N}
As a random sequence in $U(\rH_0)$, the solution of the equation 
$$
V_{n+1}=\sum_{k=0}^N B_k V_n X^k_{n+1}
$$
is an homogeneous Markov chain on $U(N)$ (actually a standard random
walk), described as follows:
$
V_0=I
$ almost surely 
and $V_{n+1}$ takes one of the values $W_iV_n$, $i\in\{0,1,...,N\}$,
with respective probability $p_i$, independently of $V_n$.
\end{theo}

\begin{proof}
Assume $V_n$ is given, depending on the random variables $X_1, \ldots,
X_n$ only. Then the random variable $X_{n+1}$ is independent and
$X^i_{n+1}=v_l^i$, with probability $p_l$.  Therefore, with
probability $p_l$ we get
$$V_{n+1}=\sum_{i=0}^N\,B_iv_l^i\,V_n=W_lV_n,\,.$$
This proves the result.
\end{proof}

\section{The Case $N=1$}\label{N=1}

In order to illustrate the results of the previous section, we detail here
the situation in the case $N=1$. 

\smallskip
Consider the set $\Omega=\{0,\,1\}^\NN$, equiped with the
$\sigma$-field $\mathcal{F}$ generated by finite cylinders. We denote 
by $\nu_n$ the coordinate mappings, for all $n\in\NN$, that is $\nu_n(\omega)=\omega(n)$. 

For $p\in]0,1[$ and $q=1-p$, we define the probability measure $\mu_p$
on $(\Omega,\,\mathcal{F})$ which makes $(\nu_n)_{n\in\NN}$ to be a
sequence of independent, identically distributed, Bernoulli random
variables with law $p\delta_1+q\delta_0$. We denote by
$\mathbb{E}_p$ the expectation with respect to $\mu_p$.

Define the random variables
$$X_n=\frac{\nu_n-p}{\sqrt{pq}}\,.$$
They satisfy $\mathbb{E}_p[X_n]=0$ and $\mathbb{E}_p[X_n^2]=1$, hence
they are obtuse random variables in $\RR$. They  take the two values
$v_0=\sqrt{q/p}$ and $v_1=-\sqrt{p/q}$ with respective probabilities
$p$ and $q$. 

The  3-tensor $T$ associated to $X$ is easy to determine. Indeed, one
can easily check the following multiplication formula.
\begin{prop}
We have
$$X_n^2=1+c_pX_n,$$
where $c_p=\frac{q-p}{\sqrt{pq}}.$
\end{prop}
This means that the 3-tensor in this context, which is a constant, is $T=c_p$.

\smallskip
In this context also, note that the space $T\Phi(X)$ is the space
$L^2(\Omega,\,\mathcal{F},\,\mu_p)$, whereas the space $\TF$ is
$\otimes_{i\in\NN}\,\CC^2$.
As an application of Theorem \ref{identi}, the operator of
multiplication by $X_n$ on $T\Phi(X)$ is represented on $\TF$ as 
$$
M_{X_n}^p=a^0_1(n)+a^1_0(n)+c_pa^1_1(n)\,.
$$

Here we are, we have put all the corresponding notations. We can apply
Theorem \ref{iff} to this particular case.

\begin{theo}
Consider the obtuse random walk ${(X_n)}_{n\in\NN}$ on $\RR$, as
described above.
Then the repeated quantum interaction evolution equation
$$
V_{n+1}=\sum_{i,j=0}^N U^i_jV_n\, a^i_j(n+1)
$$
takes the form
$$
V_{n+1}=B_0 V_n + B_1V_n X_{n+1}
$$
if and only if there exist 2 unitary operators $W_0$ and $W_1$ on
$\rH_0$ such that
$$
U=\left(\begin{matrix}pW_0+qW_1&\sqrt{pq}\,(W_0-W_1)\\
\vphantom{\sum_{i=1}^N}\sqrt{pq}\,(W_0-W_1)&qW_0+pW_1\end{matrix}\right) \,.
$$
In that case, the coefficients $B_i$ above are given by
$$
B_0=U_0^0\,,\ \ \ B_1=U^0_1=U_0^1\,.
$$
The random sequence ${(V_n)}_{n\in\NN}$ is defined by $V_0=I$ and 
$$
V_{n+1}=\begin{cases} W_0\,V_n&\mbox{\ with probability } p\\
W_1\,V_n&\mbox{\ with probability } q\,.
\end{cases}
$$
\end{theo}

\end{document}